\documentclass[Afour,sagev,times]{sagej}
\usepackage{amsmath,amsfonts}
\usepackage{algorithmic}
\usepackage{algorithm}
\usepackage{array}
\usepackage[caption=false,font=normalsize,labelfont=sf,textfont=sf]{subfig}
\usepackage{textcomp}
\usepackage{stfloats}
\usepackage{url}
\usepackage{verbatim}
\usepackage{graphicx}
\usepackage{multirow}
\usepackage{booktabs}
\usepackage{hyperref}
\usepackage{siunitx}
\usepackage{setspace}

\usepackage[nocfg]{nomencl}
\makenomenclature
\usepackage{etoolbox}
\renewcommand\nomgroup[1]{%
	\item[\bfseries
	\ifstrequal{#1}{A}{Abbreviations}{%
		\ifstrequal{#1}{V}{Variables}}%
	]}
\newcommand\BibTeX{{\rmfamily B\kern-.05em \textsc{i\kern-.025em b}\kern-.08em
		T\kern-.1667em\lower.7ex\hbox{E}\kern-.125emX}}

\def\volumeyear{2024}

\begin{document}

\title{A Review of Carsickness Mitigation: Navigating Challenges and Exploiting Opportunities in the Era of Intelligent Vehicles}

\author{Daofei Li*, Tingzhe Yu, Binbin Tang}

\affiliation{Institute of Power Machinery and Vehicular Engineering, Zhejiang University, China}

\corrauth{Daofei Li, Institute of Power Machinery and Vehicular Engineering, Zhejiang University, Hangzhou, 310027, China}

\email{dfli@zju.edu.cn}

\begin{abstract}
Motion sickness (MS) has long been a common complaint in road transportation. However, in the era of driving automation, MS has become an increasingly significant issue. The future intelligent vehicle is envisioned as a mobile space for work or entertainment, but unfortunately passengers' engagement in non-driving tasks may exacerbate MS. Finding effective MS countermeasures is crucial to ensure a pleasant passenger experience. Nevertheless, due to the complex mechanism of MS, there are numerous challenges in mitigating it, hindering the development of practical countermeasures.
To address this, we first review two prevalent theories explaining the mechanism of MS. Subsequently, this paper provides a summary of current subjective and objective approaches for quantifying motion sickness levels. Then, it surveys existing methods for alleviating MS, including passenger adjustment, intelligent vehicle solutions, and motion cues of various modalities. Furthermore, we outline the limitations and remaining challenges of current research and highlight novel opportunities in the context of intelligent vehicles. Finally, we propose an integrated framework for alleviating MS. The findings of this review will enhance our understanding of carsickness and offer valuable insights for future research and practice in MS mitigation within modern vehicles.
\end{abstract}


\maketitle

\section{Introduction}
In recent years, the intelligent vehicle has seen rapid development and significant advancements, integrating various technologies such as environment perception, automated decision and control, on-board computing, intelligent cockpit, and vehicle-to-everything connectivity. The rise of conditionally automated or fully autonomous driving is expected to become a reality in the near future. As a result, road transportation users naturally anticipate safer, more efficient, and more comfortable travel experiences.

With driving automation, drivers are partially or fully relieved from regular driving tasks, allowing them to transition into the role of passengers. This means that all on-board passengers, including former drivers, can maximize their travel time by engaging in other activities such as meetings, reading, watching movies, gaming, and more.
Unfortunately, such disengagement from driving tasks and subsequent role change to passengers raise concerns about motion sickness (MS). 
A reasonable explanation for this phenomenon is the so-called ``driver-passenger effect", which suggests that passengers not controlling the vehicle are more susceptible to MS compared to the driver of the same vehicle \cite{Rolnick_1991_whydriverrarely}.
To make it worse, the rapid commercialization of electric vehicles has introduced new carsickness challenges due to their distinct vehicle dynamics compared to that of traditional fueled cars. 

MS, a time-worn complaint in transportation, is marked by the onset of nausea and discomfort.
Severe MS can lead to a range of symptoms, including nausea, yawning, paleness, sweating, stomach awareness, increased saliva production, hiccups, headaches, blurred vision, dizziness, drowsiness, spatial orientation difficulties, difficulty concentrating, and vomiting \cite{Leung_2019_motionsicknessoverview}.
This issue is prevalent worldwide, as evidenced by a survey revealing that 46\% of participants from Brazil, China, Germany, the United States, and the United Kingdom experienced some degree of MS while traveling \cite{Schmidt_2020_internationalsurveyincidence}.

Given the significant impact of MS on the user ride experience, it has been a focal point for researchers in recent decades. Despite this attention, addressing MS in road vehicles still presents a considerable challenge. It is somewhat ironic that this age-old issue persists even as modern, electric, intelligent, and autonomous vehicles continue to advance. However, considering the recent progress in driving automation and intelligent human-vehicle interaction, there may be new opportunities to explore.

To effectively address these challenges, it is essential to systematically review existing research, identify gaps, and uncover potential opportunities for alleviating MS in intelligent vehicles. This paper aims to achieve this by reviewing research on MS in intelligent vehicles from three key perspectives: 1) theories of MS mechanism, 2) quantification of MS, and 3) countermeasures for MS. Building on the identified research gaps and future directions, we also present an example framework for mitigating carsickness.

The rest of the paper is organized as follows.
Firstly, two popular theories of MS mechanism are reviewed in \autoref{sec:TheoryMS}.
Then in \autoref{sec:quantifyMS} we discuss how to quantify MS susceptibility and MS level, including various subjective questionnaires and objective indicators of MS that are adopted in the literature. 
Next, in \autoref{sec:CounterMS}, three categories of MS countermeasures are summarized, i.e. what passengers can do (\autoref{subsec:passengerAdjust}), what vehicles can do (\autoref{subsec:intellVeh}), and how to prompt passengers with motion cues (\autoref{subsec:motioncues}).
Afterwards, in \autoref{sec:discuss}, the remaining challenges of research on carsickness are discussed, while future opportunities in tackling MS on vehicles are prospected.
Highlighting the emerging opportunities with intelligent vehicles, an integrated framework for mitigating MS is proposed and detailed in \autoref{sec:framework}.
Finally, \autoref{sec:conclude} concludes the paper.

\section{Theories of motion sickness}
\label{sec:TheoryMS}
\subsection{Sensory conflict theory}

Sensory conflict theory proposed by Reason and Brand, also known as sensory rearrangement or neural mismatch hypothesis, is one of the most popular theories about MS \cite{Reason_1978_motionsicknessadaptation}. This theory suggests that MS occurs when the perceived motion state from the vestibular system, proprioceptors, and visual receptors conflicts with internal empirical model.

Based on sensory conflict theory, a more detailed heuristic mathematical model was developed, which introduced observer theory from control engineering to describe the dynamic coupling between the putative conflict signals and nausea magnitude estimates \cite{Oman_1990_motionsicknesssynthesis}.

In order to clearly quantify the relationship between sensory conflict and MS severity, this model has to be simplified further. Thus, Bles et al. redefined the classic sensory conflict theory and presented the subjective vertical conflict (SVC) model, which simplifies sensory conflict to the single vertical dimension \cite{Bles_1998_motionsicknessonly}. All situations that provoke MS are characterized by a condition in which the sensed vertical as determined on the basis of integrated information from eyes, vestibular system and nonvestibular proprioceptors is at variance with the subjective vertical as expected from previous experience. In the case of passive vertical motion, SVC model can predict successfully relationship between MS severity and vertical sensory conflict when head acceleration in vertical direction is used as input \cite{Bos_1998_modellingmotionsickness}.

Undoubtedly, the occurrence of MS is not just a result of sensory conflict in vertical direction, e.g. occupants may suffer from horizontal stimuli when riding a car. Therefore, a six-degree-of-freedom subjective vertical conflict (6DOF-SVC) model was extended based on SVC model, which can take both linear and rotational motions of passenger head into account to estimate motion sickness incidence (MSI) \cite{NorimasaKamiji_2007_modelingvalidationcarsickness,Wada_2015_mathematicalmodelmotion}.

\subsection{Postural instability theory}

Another perspective on MS is the postural instability theory \cite{Riccio_1991_ecologicaltheorymotion}, which posits that animals become sick in situations where they cannot maintain postural stability. Further, it was found that postural sway increases before the onset of MS symptoms, which suggests that postural instability can work as a key prediction of MS \cite{Stoffregen_1998_posturalinstabilityprecedes}. 
However, there are some opposition views on this theory. For example, some researches pointed out that MS and postural instability may be both second-order effects under control of a common center \cite{Kennedy_1996_posturalinstabilityinduced}. Another research even held a view that postural instability is neither a necessary nor sufficient condition for MS \cite{Bos_2011_nuancingrelationshipmotion}. 

\subsection{Summary}
These two theories attempt to interpret how MS occurs, while there are also some other hypotheses on why MS occurs, e.g. the evolutionary hypothesis of MS \cite{Treisman_1977_motionsicknessevolutionary}, or the negative reinforcement MS model \cite{Bowins_2010_motionsicknessnegative}. 
The sensory conflict and postural instability theories are the most popular but still being debated, of which the main reason is that they are based on two fundamentally-different epistemologies \cite{Stoffregen_1998_posturalinstabilityprecedes}. However, as the later \autoref{sec:CounterMS} shows, both of them can provide insights on how to mitigate carsickness. 


\section{Quantifying motion sickness}
\label{sec:quantifyMS}
\begin{figure}[htbp]
	\centering
	\includegraphics[width=3in]{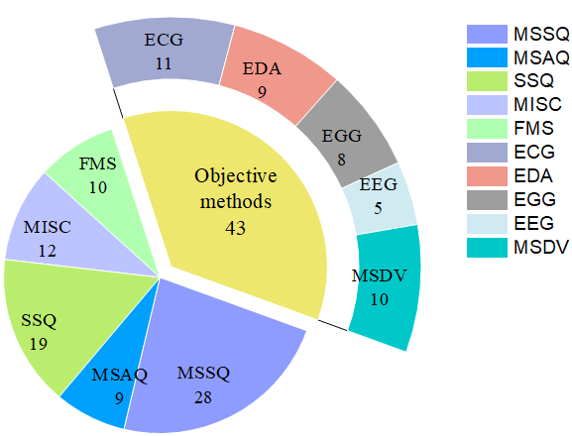}
	\caption{Statistics of MS quantification methods (number of use count).}
	\label{fig_MSquantificationpiechart}
\end{figure}

\begin{figure*}[htbp]
	\centering
	\includegraphics[width=1.0\textwidth]{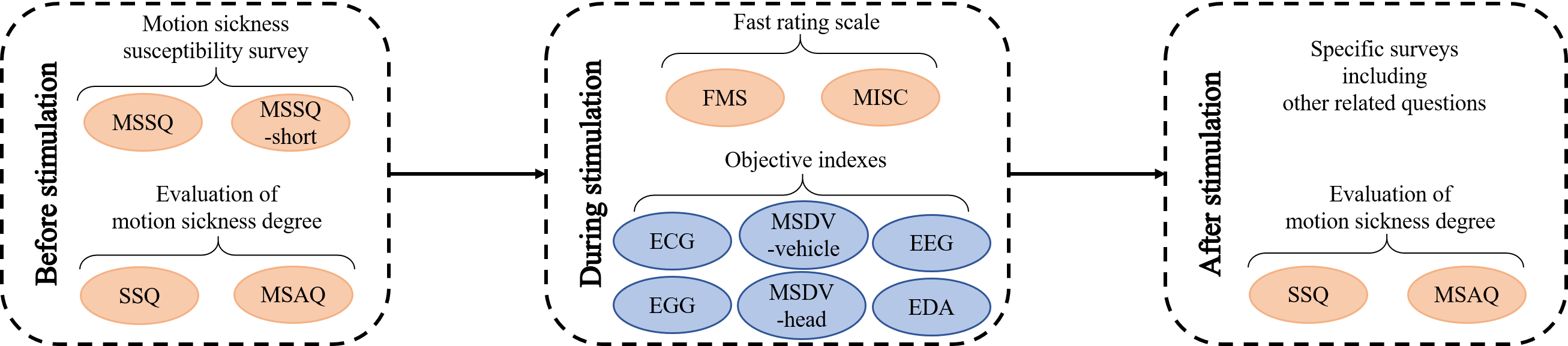}
	\caption{MS quantification in experiment: MS susceptibility and sickness level.}
	\label{fig_MSquantificationframework}
\end{figure*}

MS can be quantified using subjective and objective methods. 
Fig. \ref{fig_MSquantificationpiechart} shows the use count of various MS quantification methods in the articles reviewed here.
It can be seen that subjective questionnaire methods are more popular in current literature, accounting for more than half, though physiological indicators are also adopted to describe MS severity. 
Note that in a considerable amount of research multiple methods are often combined to reliably record the progressing of subjects' MS, as shown in Fig. \ref{fig_MSquantificationframework}. 

\subsection{Subjective methods}
Subjective questionnaires are aimed at quantifying MS susceptibility or MS status, which are summarized in Table \ref{tab:table1}.

\begin{table*}[htbp]
	\caption{Popular MS subjective questionnaires\label{tab:table1}}
	\centering
	\resizebox{\textwidth}{!}{
		\begin{tabular}{llll}
			\toprule
			~ & Reference (author, year) & Subjective method & Feature \\ \midrule
			~ & ~ & ~ & ~ \\ 
			~ & \multirow{2}{*}{Golding, 1998\cite{Golding_1998_motionsicknesssusceptibility}} & \multirow{2}{*}{MSSQ} & Investigation of MS history \\ 
			& ~ & ~ & Pre assessment of MS susceptibility \\ 
			MS &&&\\
			history questionnaire & Golding, 2006\cite{Golding_2006_predictingindividualdifferences} & MSSQ-short & Simplified version of MSSQ \\ 
			&&&\\
			~ & \multirow{2}{*}{Keshavarz et al., 2023\cite{Keshavarz_2023_visuallyinducedmotion}} & \multirow{2}{*}{VIMSSQ} & Based on MSSQ \\ 
			~ & ~ & ~ & Exclusively for visually induced MS \\ \midrule
			~ & ~ & ~ & ~ \\ 
			& \multirow{2}{*}{Kennedy et al., 1993\cite{Kennedy_1993_simulatorsicknessquestionnaire}} & \multirow{2}{*}{SSQ} & Detailed assessment \\ 
			& ~ & ~ & Three dimensions (nausea, oculomotor, disorientation) \\ 
			&&&\\
			MS& \multirow{3}{*}{Gianaros et al., 2001\cite{Gianaros_2001_questionnaireassessmentmultiple}} & \multirow{3}{*}{MSAQ} & Detailed assessment \\ 
			multidimensional questionnaire & ~ & ~ & Four dimensions \\ 
			&&&(gastrointestinal, central, peripheral, sopite-related) \\
			&&&\\
			~ & \multirow{3}{*}{Kim et al., 2018\cite{Kim_2018_virtualrealitysickness}} & \multirow{3}{*}{VRSQ} & Simplified version of SSQ \\ 
			~ & ~ & ~ & Only retain two dimensions more relevant to VR \\ 
			&&&(oculomotor and disorientation) \\\midrule
			
			~ & ~ & ~ & ~ \\ 
			~ & \multirow{3}{*}{Bos et al., 2005\cite{Bos_2005_motionsicknesssymptoms}} & \multirow{3}{*}{MISC} & 0-10 points \\ 
			~ & ~ & ~ & Fast assessment \\ 
			& ~ & ~ & Single dimension \\ 
			MS &&&\\
			fast scoring scale & \multirow{3}{*}{Keshavarz and Hecht, 2011\cite{Keshavarz_2011_validatingefficientmethod}} & \multirow{3}{*}{FMS} & 0-20 points \\ 
			~ & ~ & ~ & Fast assessment \\ 
			~ & ~ & ~ & Single dimension \\ 
			\bottomrule
	\end{tabular}}
\end{table*}

\subsubsection{Quantification of MS susceptibility}

Several MS history questionnaires, which measure MS susceptibility based on the interviewee's previous experience of MS rather than quantify interviewees' MS status in real time, are often used to rank and screen interviewees before experiment.

Motion Sickness Susceptibility Questionnaire (MSSQ) proposed by Golding \cite{Golding_1998_motionsicknesssusceptibility} investigates a subject's personal MS experience when riding in different transportation and entertainment facilities during both childhood and adulthood, which can provide an assessment of one's susceptibility to MS. 
A more simplified MSSQ-short was revised for faster evaluation while ensuring reliability \cite{Golding_2006_predictingindividualdifferences}. 

However, either MSSQ versions still have limitations in accurately assessing one's MS susceptibility, since some subjects with very low total scores may still experience severe MS in practice. This contradiction has been observed in some studies \cite{Lamb_2015_mssqshortnormsmaya,Li_2023_mosiappmotion}. This might be due to that MSSQ and MSSQ-short are made up of two components, i.e. MS experiences in the childhood (before the age of 12) and in the last decade, respectively. Some subjects have little experience of MS or even little experience of transport and entertainment facilities during their childhood, but they do have experienced MS in the last decade, which results in a relatively low total score. Therefore, it is unprecise
to judge whether one person is susceptible to MS solely based on their total score of MSSQ. 
A recommended approach in practice is to focus more on the recent MS experiences, but not on that in the childhood. Further, those items relevant to MS causes in a particular research to be conducted, such as MS on a moving bus, are more informative for screening subjects.

Recently, with the increasingly popular usage of visual devices, such as smartphones, TVs and head-mounted displays, Visually Induced Motion Sickness (VIMS) becomes common. As VIMS is different from the usual MS induced by motion stimuli, a customized questionnaire is required. For this, VIMS Susceptibility Questionnaire (VIMSSQ) was proposed based on MSSQ \cite{Keshavarz_2023_visuallyinducedmotion}. Unlike MSSQ, VIMSSQ investigates how often interviewees use various visual devices and their corresponding experience of MS, rather than that of using different transportation and entertainment facilities.

\subsubsection{Quantification of MS status}

Another version of MS assessment questionnaire was developed to quantify the subject's MS status, like Simulator Sickness Questionnaire (SSQ) \cite{Kennedy_1993_simulatorsicknessquestionnaire}. SSQ is almost indispensable for any simulator experiments on MS, due to its comprehensiveness and trustworthiness. 
To detail, SSQ consists of 16 scores for MS symptoms and 4 subscales with different dimensions, i.e. oculomotor, disorientation, nausea and total score. 
Users can obtain more specific and detailed references from subscales, which has been proven to be more related to MS compared to the total SSQ score \cite{Balk_2013_simulatorsicknessquestionnaire}. 
Similarly, Motion Sickness Assessment Questionnaire (MSAQ) is also constructed from 4 dimensions, i.e. gastrointestinal, central, peripheral and sopite-related, rather than just quantifying MS as a unidimensional construct \cite{Gianaros_2001_questionnaireassessmentmultiple}. Nevertheless, SSQ may not be appropriate for some novel virtual environments, such as virtual reality, since there were significant differences in terms of SSQ scores and the order of contributions to total sickness in the VR environment compared to that in simulators. Therefore, to generate a new questionnaire that can measure VR-induced MS more efficiently, SSQ items not relevant to VR environment should be removed \cite{Kim_2018_virtualrealitysickness}.

With either SSQ or MSAQ, interviewees have to take at least several minutes to finish their multiple dimensional questions, which means that their MS performances cannot be recorded in real time. Even if applied in a continuous process, interviewees must pause for a few minutes to complete these subjective questionnaires.
To enable real-time assessment, simpler and faster scoring scales have emerged. Among them, Misery Scale (MISC), a scoring scale with increasing levels of MS from 0 to 10, can easily be applied repeatedly since its rating takes only a few seconds \cite{Bos_2005_motionsicknesssymptoms}. Although MISC includes symptoms other than nausea, such as dizziness, headache, sweating, etc., this scale may be not easy for interviewees to distinguish these symptoms well. Similarly, Fast MS Scale (FMS) can also record MS and capture its time course fast and effectively during experiments \cite{Keshavarz_2011_validatingefficientmethod}. Both FMS and MISC have one common drawback that they inevitably sacrifice their comprehensiveness for usage convenience. In other words, these two scales may reach their limits when particular physiological symptoms become the focus of interest, owing to their inability to distinguish between feelings of nausea and its precursors, such as drowsiness.

\subsection{Objective methods}
Objective methods are usually used as supplements to subjective methods, including electrogastrography (EGG), electrodermal activity (EDA), electroencephalography (EEG), electrocardiography (ECG), etc.
Motion sickness dose value (MSDV), though slightly different, can also be categorized as an objective method, since it is derived from the accelerations exerted on interviewee, or directly from vehicle accelerations in some research.

\subsubsection{MSDV}
MSDV was introduced as an objective indicator of MS dose in ISO 2631-1:1997 standard, ``Mechanical vibration and shock - Evaluation of human exposure to whole-body vibration - Part 1: General requirements" \cite{ISO_1997_mechanical}. 
MSDV is defined as
\begin{equation}
	\label{equation1}
	MSDV=\sqrt{\int_0^T[\tilde{a}(t)]^2 d t}
\end{equation}
where $\tilde{a}$ is the frequency-weighted acceleration, and $T$ is the total exposure time under such acceleration. $\tilde{a}$ is calculated according to
\begin{equation}
	\label{equation2}
	\tilde{a}=a \times W_f
\end{equation}
where $W_f$ is the weighting factor per ISO 2631-1:1997.

Although MSDV was initially used for vertical motion, it had been proven suitable for predicting MS evoked by horizontal motion in road vehicles \cite{Turner_1999_motionsicknesspublic}. 
Due to its easiness in usage, MSDV is quite popular as an objective indicator for simulation validation. Particularly in MS researches using only simulation for validation, subjective feedback and objective physiological data cannot be collected, while MSDV from vehicle acceleration may be used as the only MS indicator \cite{Htike_2020_minimisationmotionsickness,Htike_2022_fundamentalsmotionplanning,Li_2021_mitigatingmotionsickness,Zheng_2023_mitigatingmotionsickness}. 

For human-in-the-loop experiments, MSDV can be obtained by using an accelerometer fixed to vehicle or simulator platform. If available, the passenger's head accelerations are more suitable to calculate their MSDVs, as it is the direct cause of occupant MS. However, it should be noted that passengers' active head motion, if not limited, may impact MSDV and must be considered in detailed analysis \cite{Hong_2022_adaptivecruisecontrola,Karjanto_2021_onroadstudymitigating}.

\subsubsection{EGG}
EGG has been used quite early on to characterize MS, since MS is almost always accompanied with nausea caused by tachygastria, while EGG is an accurate and noninvasive way to detect tachygastria. A close correlation has been discovered over time between tachygastria and MS symptoms \cite{Stern_1987_spectralanalysistachygastria}.
When subjects are exposed to an optokinetic rotating drum, the development of MS is accompanied by an increase in tachygastria \cite{Hu_1991_motionsicknessseverity}. Further, it was found that the increased EGG activity at 4-9 cycles per min can indicate the severity of MS \cite{Hu_1999_systematicinvestigationphysiological}. A recent study also confirmed that reported nausea was accompanied by the increase in amplitude and root mean square value of EGG during the driving simulation \cite{Gruden_2021_electrogastrographyautonomousvehicles}. 
However, there is still debate about whether EGG is reliable and robust enough to indicate MS, since the inherent variability of EGG and inter-individual variability are unavoidable \cite{Cheung_1998_perspectiveselectrogastrographymotion}.

\subsubsection{ECG}

ECG is also often utilized to evaluate MS. One ECG indicator used frequently is heart rate (HR), which was found to change significantly during the nauseogenic rotating chair test \cite{Cowings_1990_stabilityindividualpatterns}. Specifically, HR would increase significantly with increasing subjective ratings of MS \cite{Holmes_2001_correlationheartrate}. Except for HR, Heart rate variability (HRV) also draws much attention. For example, for those subjects susceptible to MS, the power spectrum density of R-R interval was monitored to reduce significantly at the mid and high frequencies during a brief vestibular disorientation test \cite{Doweck_1997_alterationsvariabilityassociated}. 
The enhancement of LF:HF ratio, where LF means the low-frequency power of HRV and HF means the high-frequency power of HRV, may also indicate more severe symptoms of MS \cite{Yokota_2005_motionsicknesssusceptibility}. 
On the contrary, a recent research detected that the LF:HF ratio still declined as the MS stimulation progressed \cite{Irmak_2021_objectivesubjectiveresponses}. For this contradictory phenomenon, one feasible explanation is that LF:HF ratio is actually an integrated reflection of sympathetic and parasympathetic activity rather than a simple linear measure of sympathovagal balance as previously thought.

\subsubsection{EDA}
When suffering from MS, sweating usually intensifies, thus the skin conductance increases and skin resistances decreases. With this, it is possible to monitor MS using EDA. Note that the sensitivity of electrical activity to MS varies among different body parts, while areas with well-developed sweat glands are used commonly for detecting EDA, such as fingers, palms and feet. Compared to that at fingers and palms, the phasic skin-conductance responses recorded at the forehead site are more sensitive physiological correlates of MS \cite{Golding_1992_phasicskinconductance,Wan_2003_correlationphasictonic}. One recent research also confirmed that forehead humidity would increase significantly when passengers experiencing MS during actual driving \cite{Bando_2021_developmentevaluatingmethods}. However, it is possible that EDA may change over time even without MS \cite{Irmak_2021_objectivesubjectiveresponses}. Similar to other physiological indicators, EDA also has significant inter-individual variability due to various factors such as environment, emotional state and sweat gland development \cite{Smyth_2021_exploringutilityeda}. 

\subsubsection{EEG}
With the development of machine learning techniques, EEG has the potential to be the most reliable predictor of MS. Researchers have tried to link brain signals to symptoms of MS. 
For example, a phenomenon was found that EEG power spectral levels in the delta and theta bands increased along with the level of MS \cite{Chelen_1993_spectralanalysiselectroencephalographic}. As technology advances, the power signals of five brain areas, i.e. the left motor, the parietal, the right motor, the occipital and the occipital midline, were found highly correlated with MS level \cite{Chen_2010_spatialtemporaleeg,Li_2022_eegbasedevaluationmotion}. In particular, the successful introduction of machine learning methods into EEG analysis enables that EEG features most associated with MS can be extracted. Utilizing models based on mass-extracted EEG features, the accuracy of predicting MS symptoms can reach over 80\% \cite{Chin-TengLin_2013_eegbasedlearningsystem,Liao_2020_usingeegdeep}. Optimistically speaking, EEG is expected to become one of the most reliable ways to quantify MS, if supported with sufficient MS data in the future.

\subsubsection{Other}
Last but not least, the postural sway is supposed to serve as reliable predictors for the reason that postural instability precedes MS \cite{Owen_1998_relationshipposturalcontrol,L.JamesSmart_2002_visuallyinducedmotion}. Further, body temperature is considered to be a possible index due to cutaneous vasodilation and sweating induced by MS, which can lead to heat loss and hypothermia \cite{Nalivaiko_2014_motionsicknessnausea}.

\subsection{Summary}
\begin{enumerate}
	\item \textbf{Proper methods during experiment}. The two categories of MS quantification methods are usually combined used all through the different phases of MS experiment, as also shown in Fig. \ref{fig_MSquantificationframework}.
	All through the entire experiment, objective indicators should be collected as supplements to subjective reports.  
	Specifically, before the experiment stimulation occurs, MSSQ or MSSQ-short can be used to assess the MS susceptibility of subjects. Here, SSQ or MSAQ can be collected to evaluate the initial MS level as a baseline.
	During the experiment stimulation, FMS or MISC can be used as a fast scoring scale to sample the development of MS in the subjects. 
	After the experiment stimulation ends, SSQ or MSAQ can be used again to record the final state of the subjects. If the MS recovery process is of concern, all related indicators or subjective reports can be sampled during a certain period after stimulation.
	\item \textbf{Limitations of physiological indexes}. To our regret, whether EGG, EEG, ECG, EDA or MSDV, the clear correlation between these indexes and MS has not been clearly revealed, partly due to the fact that MS is rather complicated and there are still many mysteries regarding the underlying mechanisms. This may also further prevent successful establishment of an accurate mapping between these single indicators and MS. 
	\item \textbf{Potential of multi-modal fusion}. Although it is currently still difficult for a single index to accurately characterize MS, multi-modal data fusion approaches based on machine learning are increasingly favored. For example, a dataset containing various physiological signals from real driving scenarios can be constructed and then several features related to MS are extracted, with which the prediction model after training can achieve high accuracy \cite{Tan_2022_motionsicknessdetection,Hwang_2022_classificationmotionsickness}. 
	This indicates that the fusion of multiple indexes at different stages, as shown in Fig. \ref{fig_MSquantificationframework}, may be the optimal way to quantify individual MS level in the future.
	\item \textbf{MS quantification in real practice}. For future practical use, non-contact types of MS measures are preferred, e.g. MSDV, non-infrared or infrared camera, while wearable devices may also be used to get ECG and EDA measures. Before such measures converge to a specific passenger's MS status, data collection and training should be carried out, with the help of subjective measures (e.g. MISC). Note that this is still an underexplored area of research, which calls for extensive attempts in real vehicle tests on real traffic contexts.  
\end{enumerate}

\section{Countermeasures for motion sickness}
\label{sec:CounterMS}
\begin{figure*}[htbp]
	\centering
	\includegraphics[width=1\linewidth]{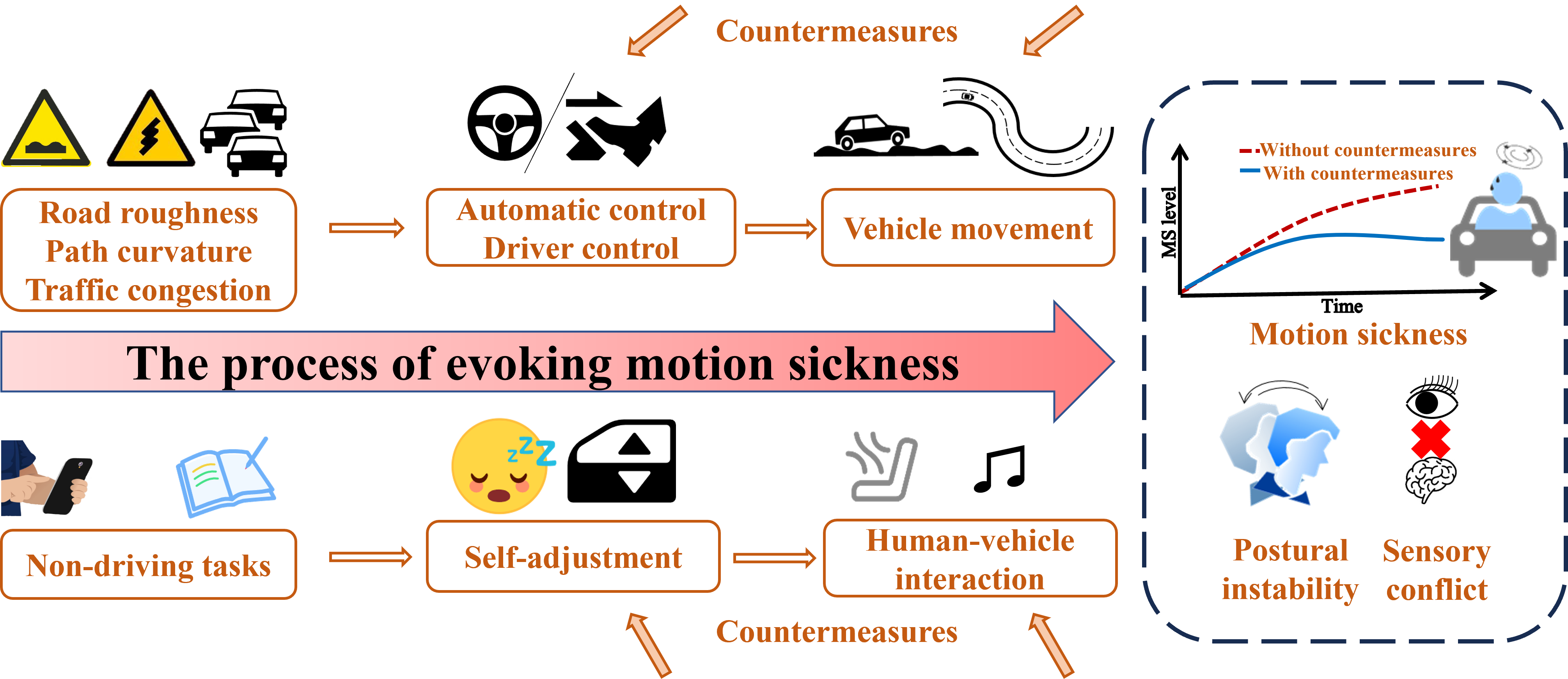}
	\caption{The evoking and mitigating of motion sickness (upper: stimulation side, lower: human side).}
	\label{fig_MSevokeandmitigate}
\end{figure*}

Recalling \autoref{sec:TheoryMS}, it is clear that countermeasures for MS should consider both the human side and stimulation side. 
Corresponding to the two sides, Fig. \ref{fig_MSevokeandmitigate} illustrates the MS evoking process and potential on-vehicle alleviation solutions. 

For the stimulation side, congested traffic, path curvature and road roughness, along with the vehicle control, are the root causes to induce non-smooth movements and thus MS. 
For the human side, passenger MS is further exacerbated by inappropriately allocating visual attention, e.g. engaging non-driving tasks.  
Note that specific anti-MS medicines, though not detailed in this survey, can be taken before, during or after car rides, for which readers may refer to a recent review paper \cite{Koch_2018_neurophysiologytreatmentmotion}. 
Referring to either sensory conflict or postural instability theories, the human's MS status actually depends on the meeting point of human and stimulation, i.e. how the human responds to the stimulation, both mentally and physically. 
Such human responses may include the active adjustment of cabin environment, or re-allocation of attention for better comprehension and anticipation of vehicle movements.  
With this in mind, here the existing research on MS countermeasures are categorized into the following three kinds, i.e. passengers, vehicles and motion cues.

\subsection{Passengers' active adjustment}\label{subsec:passengerAdjust}
Aside from medical relief, one of the most famous and effective methods is habituation training for MS alleviation. For this, subjects would be repeatedly placed in a provocative environment for a long time in order to adapt to the stimulus. Almost all individuals who have undergone habituation training can gain some resistance to MS. Habituation training has been widely used in military, especially in the MS desensitization program for pilots, since for them the side effects such as drowsiness and fatigue caused by anti-MS drugs are unacceptable. 
Being exposed continuously to the provocative environment, people can gradually adapt and eventually gain stable MS desensitization without side effects \cite{McCauley_1976_motionsicknessincidence}. In particular, repeated exposure to multiple cross-coupled stimuli, such as a combined visual-vestibular stimulation, rather than just a single stimulus, can result in better desensitization \cite{Dai_2011_prolongedreductionmotion}.
However, though with no side effects of anti-MS drugs, there comes more cost of additional long-term training.

On the other hand, there are also other countermeasures that passengers can take, as shown in Table \ref{tab:table2}. According to sensory conflict theory, the habituation approach is fundamentally trying to help human adapt to the external stimuli by changing the internal model. Or we may also sleep or just doze with eyes closed for visual occlusion \cite{Ishak_2018_visualocclusiondecreases}. Then the visual input from eyes would be eliminated by this method, thereby reducing sensory conflict. Likewise, empirical studies have proven that looking out the window to obtain a stable external horizon reference helps eliminate passenger discomfort \cite{Turner_1999_motionsicknesspublic,Bos_2005_motionsicknesssymptoms}. Alternatively, passengers can align head and body with gravitoinertial force deliberately to reduce the perceived motion conflict \cite{Golding_2003_motionsicknesstilts}. In other words, MS symptoms can be alleviated through lying supine or tilting the head in the centripetal direction when there is a longitudinal acceleration or centripetal acceleration \cite{Wada_2012_canpassengersactive,Wada_2018_analysisdriverhead}. 

In fact, many other relaxing methods may work for alleviating MS, yet their effectiveness still remains controversial. For instance, listening to pleasant music has been found beneficial in the relief of MS \cite{Sang_2006_behavioralmethodsalleviating,Keshavarz_2014_pleasantmusiccountermeasure}. Moreover, just as what we do when we are nervous, slow and deep breathing or chewing gum may inhibit the discomfort caused by MS \cite{Jokerst_1999_slowdeepbreathing,Kaufeld_2022_chewinggumreduces}. Actually any means of mental distraction may have a positive impact on passengers suffering from MS, since successful mental distraction may relieve some pain \cite{Bos_2015_lesssicknessmore}. In addition to these methods, passengers can even control the diet to avoid over-eating or take anti-MS drugs in advance \cite{Koch_2018_neurophysiologytreatmentmotion}. 
Further, it is reported that just positive verbal and psychological instructions are suggested due to their powerful placebo effects \cite{Horing_2013_reductionmotionsickness}.

In general, passengers' active adjustment is convenient and works immediately for daily travel. But it is still required to combine countermeasures from other aspects to eradicate MS as much as possible.

\begin{table*}[b]
	\caption{Methods for passengers to alleviate MS\label{tab:table2}}
	\centering
	\resizebox{\textwidth}{!}{
		\begin{tabular}{llll}
			\toprule
			Reference (author, year) & Experiment subjects & Analysis methods & Countermeasures\\
			\midrule
			\multirow{2}{*}{Jokerst et al., 1999\cite{Jokerst_1999_slowdeepbreathing}}&46 persons&	\multirow{2}{*}{EGG}&\multirow{2}{*}{Slow deep breathing}\\
			&17-26 years old&&\\\midrule
			
			\multirow{2}{*}{Sang et al., 2006\cite{Sang_2006_behavioralmethodsalleviating}}&24 persons (10 males, 14 females)&MSSQ&Listening to pleasant music\\
			&Average 27 years old& 1-4 points rating scale&Controling breathing\\\midrule
			
			\multirow{2}{*}{Wada et al., 2012\cite{Wada_2012_canpassengersactive}}&10 persons (9 males, 1 female)&MSSQ &Tilting the head\\
			&Average 21.5 years old&1-6 points rating scale&towards the centripetal direction\\\midrule
			
			\multirow{3}{*}{Horing et al., 2013\cite{Horing_2013_reductionmotionsickness}}&\multirow{2}{*}{32 persons (16 males, 16 females)}&MSSQ, EGG&\multirow{3}{*}{Placebo effect}\\
			&\multirow{2}{*}{Average 26 years old}&0-5 points rating scale&\\
			&&contained seven symptoms&\\\midrule
			
			\multirow{2}{*}{Bos, 2015\cite{Bos_2015_lesssicknessmore}}&16 persons (8 males, 8 females)&\multirow{2}{*}{MSSQ, MISC}&\multirow{2}{*}{Mental distraction}\\
			&Average 31.4 years old&&\\\midrule
			
			\multirow{2}{*}{Ishak, 2018\cite{Ishak_2018_visualocclusiondecreases}}&11 persons (4 males, 7 females)&\multirow{2}{*}{SSQ}&\multirow{2}{*}{Visual occlusion}\\
			&Average 27.7 years old&&\\\midrule
			
			\multirow{2}{*}{Kaufeld et al., 2022\cite{Kaufeld_2022_chewinggumreduces}}&77 persons (34 males, 43 females)&\multirow{2}{*}{MSSQ, SSQ, FMS}&Chewing gum\\
			&Average 34.01 years old& &(A pleasant odor and mental distraction)\\
			\bottomrule
	\end{tabular}}
\end{table*}

\subsection{Intelligent vehicle solutions}
\label{subsec:intellVeh}
\subsubsection{(1) Smart Cockpit}

Smart cockpit is a smart system in modern vehicles that focuses on the user experience by leveraging recent developments of seating, climate control, driver/occupant monitoring, intelligent display, and AI-assisted multi-modal interactions. As the core carrier to realize the so-called ``third living space" of automobile, it provides users with an immersive experience that combines travel, life, and entertainment, while enhancing the safety, comfort, and convenience of mobility. 

Several recommendations for cockpit-based MS mitigation have been proposed, as detailed in Table \ref{tab:table3}. For instance, a large visual field is essential for alleviating MS while maintaining vehicle safety \cite{Griffin_2004_visualfieldeffects}. The seating system can significantly impact ride comfort, too. To alleviate MS, rearward seats should be avoided since they will result in more sensory conflicts \cite{Salter_2019_motionsicknessautomated}. It is better that the seat backrest is reclined backward and has a passive restraint system such as comfortable headrests, which help to avoid passengers' excessive head movement \cite{Keshavarz_2017_passiverestraintreduces,Bohrmann_2020_reclinedpostureenabling}. The seat can be further integrated with vehicle suspension system, and its stiffness and damping adapt to road conditions to mitigate vibration stimuli, which will be covered later in the vehicle control section.

Additionally, touch screens have been extensively adopted in vehicle cockpits, while activities performed on them may also increase MS risk. As a consequence, the display size, position and content should be thoroughly optimized to minimize passengers' sickness \cite{Diels_2015_userinterfaceconsiderations,Kuiper_2018_lookingforwardinvehicle}. For example, the display should not be located lower, while small displays should be preferred.

Moreover, an intelligent vehicle cockpit is possible to implement customized countermeasures combined with those mentioned in \autoref{subsec:passengerAdjust}, such as pleasant music, odors and ventilation \cite{Keshavarz_2014_pleasantmusiccountermeasure,Keshavarz_2015_visuallyinducedmotion,DAmour_2017_efficacyairflowseat}. When the occupant monitoring system detects MS discomfort, a series of cockpit adjustments can be implemented automatically based on user preferences. For instance, it would open windows for ventilation, adjust cockpit temperature, release relaxing fragrance and so on. It should be stressed that great care has to be taken to choose a pleasant odor accepted by passengers, as some odors are probably detrimental to passengers' well-being, i.e. leading to more severe MS \cite{Schartmuller_2020_sickscentsinvestigating}.

\begin{table*}[htbp]
\caption{Intelligent vehicle cockpit solutions for MS mitigation\label{tab:table3}}
\centering
\resizebox{\textwidth}{!}{
	\begin{tabular}{llll}
			\toprule
			Reference (author, year) & Experiment subjects & Analysis methods & Countermeasures\\
			\midrule
			Keshavarz and Hecht, 2014\cite{Keshavarz_2014_pleasantmusiccountermeasure}&93 persons (43 males, 50 females)&SSQ, FMS&Pleasant music\\\midrule
			
			Keshavarz et al., 2015\cite{Keshavarz_2015_visuallyinducedmotion}&62 persons (15 males, 47 females)&SSQ, FMS&Pleasant odor\\\midrule
			
			D'Amour et al., 2017\cite{DAmour_2017_efficacyairflowseat}&82 persons (39 males, 43 females)&SSQ, FMS&Ventilation\\\midrule
			
			\multirow{4}{*}{Keshavarz et al., 2017\cite{Keshavarz_2017_passiverestraintreduces}}&21 young persons (7 males, 14 females)&\multirow{3}{*}{SSQ, FMS}&\multirow{4}{*}{Passive constraints}\\
			&Average 25 years old&\multirow{3}{*}{Postural sway}&\\
			&16 old persons (5 males, 11 females)&& \\		
			&Average 71 years old&& \\\midrule
			
			\multirow{2}{*}{Kuiper et al., 2018\cite{Kuiper_2018_lookingforwardinvehicle}}&18 persons (8 males, 10 females)&\multirow{2}{*}{MISC}&\multirow{2}{*}{Wider peripheral vision}\\
			&Average 26 years old&& \\\midrule
			
			\multirow{2}{*}{Salter et al., 2019\cite{Salter_2019_motionsicknessautomated}}&20 persons (11 males, 9 females)&SSQ&Forward facing seats\\
			&Average 36 years old&MSSQ-short&Instead of rearward facing seats\\\midrule
			
			\multirow{2}{*}{Bohrmann and Bengler, 2020\cite{Bohrmann_2020_reclinedpostureenabling}}&25 persons (21 males, 4 females)&MSSQ-short&\multirow{2}{*}{Reclining backrest}\\
			&Average 42.46 years old&MSAQ, FMS&\\
			\bottomrule
	\end{tabular}}
\end{table*}

\subsubsection{(2) Trajectory Planning}
If viewing MS as a trouble in car ride, it is better to nip it in the bud.
Considering the common-used ``sensing - prediction - planning - control" (SPPC) architecture of driving task, it is tempting to detect the potential causes of MS and then to plan a sickness-free or sickness-less trajectory for the vehicle.
In automated or semi-automated vehicles equipped with environment sensors and on-board computing, such techniques are implementable.

In human-driven vehicles, experienced drivers prefer a driving style that avoids MS since they themselves do not want to get uncomfortable, which means that human drivers can function as a type of MS predictor and moderator \cite{Wada_2016_motionsicknessautomated}. 
Unlike human-driven vehicles, autonomous vehicles are more likely to maneuver unexpectedly for passengers accustomed to manual vehicles, leading to discomfort or even MS. Empirically, passengers would not like aggressive acceleration and steering. Thus, if MS is considered in trajectory planning in advance, it can directly reduce motion stimulus and improve ride comfort. For example, in designing the transition curve for lane changing maneuvers, the 3-point B-spline was proved to be effective in preventing MS, as it is smooth enough and easy to track \cite{Siddiqi_2022_ergonomicpathplanning}.

To account for MS in trajectory planning, it is a common way to construct an optimal control problem (OCP) with the objective of minimizing MS. In other words, an index reflecting MS level is added to the cost function, and the optimal global trajectory is obtained within given constraints. 
In addition to general indexes such as safety, efficiency and comfort, the cost function of OCP taking MS into account usually includes weighted root mean square acceleration (WRMSA) \cite{ISO_1997_mechanical}, MSDV \cite{ISO_1997_mechanical} and MSI \cite{Wada_2015_mathematicalmodelmotion}. For instance, a turning trajectory was successfully generated in line with the expert driver with the objective of minimizing MSI calculated by 6DOF-SVC model \cite{Wada_2016_motionsicknessautomated}. Similarly, Htike et al.\cite{Htike_2020_minimisationmotionsickness, Htike_2022_fundamentalsmotionplanning} constructed an OCP that takes MSDV into account while adding penalties for driving too long. 
An balance between MS and travel time can be achieved by changing their weighting factors. It is obvious that MS has a time accumulation effect, rather than occurring intensively just at the beginning of journey. Hence it is not recommended to sacrifice too much travel time to alleviate MS, otherwise we may lose both traffic efficiency and ride experience. 
For this, their next experiment using human-in-loop driving simulator also proved that the reduction in MSDV value could significantly relieve passengers without sacrificing too much time efficiency \cite{Jain_2023_optimaltrajectoryplanning}. 

Nevertheless, with OCP-based framework of planning, it is worth exploring that how to find a Pareto optimal solution to meet requirements in various scenarios. Perhaps, a more satisfactory trajectory could be customized by changing weight factors of multi-objective optimization. For instance, Tang et al. ranked passengers' susceptibility to MS according to MSSQ score and customized various trajectory correspondingly, which could achieve more targeted MS mitigation effect \cite{Tang_2023_personalizedtrajectoryplanning}.

On the other hand, it is clear that accelerations at different frequencies affect MS differently, according to ISO 2631-1:1997 \cite{ISO_1997_mechanical}, and low frequency stimuli around 0.16 Hz are supposed to have the greatest impact on MS. Therefore, Li et al.\cite{Li_2021_mitigatingmotionsickness} proposed a frequency-shaping approach to optimize acceleration profile, specifically by rearranging the acceleration distribution at different frequencies to reduce low frequency acceleration around 0.16 Hz. It was shown that MSDV could be significantly reduced, since the longitudinal acceleration below 0.2 Hz was significantly lower. A similar approach was taken to reduce the acceleration distribution in the most disgusting frequency range \cite{Zheng_2023_mitigatingmotionsickness}, which was validated further by being compared to the best human driver performance on real road.

In addition to global trajectory optimization methods above, model predictive control (MPC) adopts a rolling optimization strategy and is more suitable for practical scenarios. For example, nonlinear model predictive control (NMPC) was used to optimize the speed profile on a given path, and results showed that MSI could be reduced \cite{Certosini_2019_preliminarystudymotion,Certosini_2020_optimalspeedprofile}. Another interesting phenomenon was found that MSI would be actually higher if the maximum acceleration of autonomous vehicles is limited too small. One possible explanation is that a too small maximum acceleration limit can make acceleration change last longer and more frequent, and thus longer entire journey time and less time of steady driving.


\subsubsection{(3) Vehicle Control}
With motion command from the upper-level motion planning module, the control module will regulate the traction, braking and steering systems to accelerate, decelerate and turn. In addition to tracking accuracy, passenger MS is also worth being considered comprehensively to improve ride comfort.

In the car-following scenarios of congested urban traffic, the longitudinal speed may change frequently, which is prone to evoke passenger MS. Aiming at this, Li et al. embedded 6DOF-SVC model into an MPC framework to obtain optimal following speed control strategy \cite{Li_2021_automatedcarfollowingalgorithm}. Results show that MSI and MSDV of passengers can be reduced by 7.3\% and 5.8\%, respectively. 
In another research, an MPC-based adaptive cruise control algorithm was designed by including passenger head motion in its cost function, which could reduce MSDV by 50\% under its specific condition \cite{Hong_2022_adaptivecruisecontrola}.

Lateral motion stimulus occurs when vehicles are changing lane or turning, which are both typical scenarios inducing passenger discomfort. For comfortable lane changes, Ukita et al. adopted acceleration and jerk as instantaneous comfort indexes and SVC as long-term comfort index, respectively \cite{Ukita_2020_simulationstudylanechange}. In order to further explore the relationship between lateral vehicle motion and passenger's head roll angle, a radial basis function network model was established to predict passenger's head motion \cite{Saruchi_2020_lateralcontrolstrategy}. Based on this, a fuzzy controller is then to correct the steering angle of front wheels to keep passenger's head stable.

Fundamentally, due to the inherent dynamic coupling, the longitudinal and lateral control need to be integrated to track a predetermined trajectory. For example, MPC was adopted to optimize steering, acceleration and deceleration, which could reduce MSDV and WRMSA while maintaining both longitudinal and lateral tracking performances \cite{Luciani_2020_comfortorienteddesignmodel}. Siddiqi et al. combined multiple controllers to establish an MS mitigating control system, which was more effective than a single control strategy \cite{Siddiqi_2023_motionsicknessmitigatinga}.

Aside from horizontal stimuli, vertical stimulus due to uneven road inputs should not be ignored, either. 
According to vertical dynamics, vehicle ride comfort depends on both the suspension and seating systems, i.e. their capabilities to filter out disturbing vibration transmitted from road unevenness. 
Comparing to the traditional passive suspension, semi-active or fully active suspension systems can adapt vertical body motion to road conditions for better safety and comfort, including alleviated motion sickness. 
For example, a study has shown that a high bandwidth active suspension can respond to roads at high frequency to mitigate passenger MS \cite{Ekchian_2016_highbandwidthactivesuspension}. 
Further, active control can also be integrated into vehicle seating system. 
For example, Papaioannou et al. attempted to maintain stable seat motion and then reduce MSI \cite{Papaioannou_2022_integratedactiveseat, Papaioannou_2022_kseatbasedpidcontroller}.

Active suspension systems can not only suppress vertical oscillations, but also play a role in lateral motion regulation. For instance, it was found that the impact of lateral acceleration on passengers can be effectively offset by adaptively tilting vehicle towards the center, thereby alleviating MS \cite{Zheng_2022_curvetiltingnonlinear}.
Interestingly, some other studies found that active roll stabilization and rear wheel steering systems do not have significant benefits in alleviating MS, thought certainly without negative effects \cite{Jurisch_2020_influenceactivesuspension}. A probable explanation is that MS may only be alleviated when passengers tilt actively their bodies towards the centripetal direction, rather than being passively tilted by active suspension systems \cite{Golding_2003_motionsicknesstilts}.

\subsubsection{(4) Summary}

Overall, the challenges and opportunities for researches on motion sickness-less planning and control are summarized as follow. 
\begin{enumerate}
	\item \textbf{Vehicle dynamic model}. For practical applications, vehicle models with sufficient fidelity should be used in planning. Currently, most work on planning still adopts over-simplified model of vehicle dynamics, e.g. 2 degrees of freedom model or point mass model. However, this may not work in complex driving scenarios, e.g. those with high lateral accelerations. Further, if bumpy roads are considered, integrated vehicle models including vertical dynamics should be incorporated. 
	\item \textbf{Real time planning}. Only when real-time trajectory planning becomes possible could these approaches be integrated into real vehicles. However, whether OCP or MPC-based planning methods, a huge computing power is required. With this, most existing studies can only select a specific path after offline optimization due to limitations in computing power. It is a thorny problem worth tackling the real-time performance of optimization-based planning. Along with the rapid development of automated driving, such challenges may be overcome with high-performance on-board computing in future. 
	\item \textbf{Integrated planning and control}. Most researches focus on either planning or control algorithms for MS mitigation, while for practice both algorithms should work together. With the advancements of automated driving, such integration becomes promising, but the tradeoff between MS and other objectives (e.g. safety, energy consumption) should be handled with care.
	\item \textbf{Validation approach}. Most of them were limited to simulation only, while in rare cases they were validated in driving simulators \cite{Ekchian_2016_highbandwidthactivesuspension,Zheng_2022_curvetiltingnonlinear,Jain_2023_optimaltrajectoryplanning}, 
	Currently, to our knowledge there is still no open reports in this field that uses real road test for validation.
	It is probably due to the insufficiency of vehicle model and algorithm computing efficiency.
\end{enumerate} 

%
%
%
%

\subsection{Motion Cues}\label{subsec:motioncues}
Passengers on a moving vehicle are usually not aware of the driving contexts of ego vehicle, while their lack of external visual reference often leads to MS. 
Therefore, as suggested in Fig. \ref{fig_MSevokeandmitigate}, motion cues can be provided to passengers, thus promoting passengers' anticipation or comprehension of vehicle motion. 

A non-negligible fact is that it has become common for occupants to operate non-driving tasks (NDT) on handheld or in-vehicle devices, e.g. texting or watch videos on smartphones, tablets, dashboard displays and head-mounted displays.
These devices can provide various modalities of human machine interaction, i.e. for conveying motion information via visual, haptic and auditory cues. 

\begin{figure*}[htbp]
	\centering
	\includegraphics[width=1.0\textwidth]{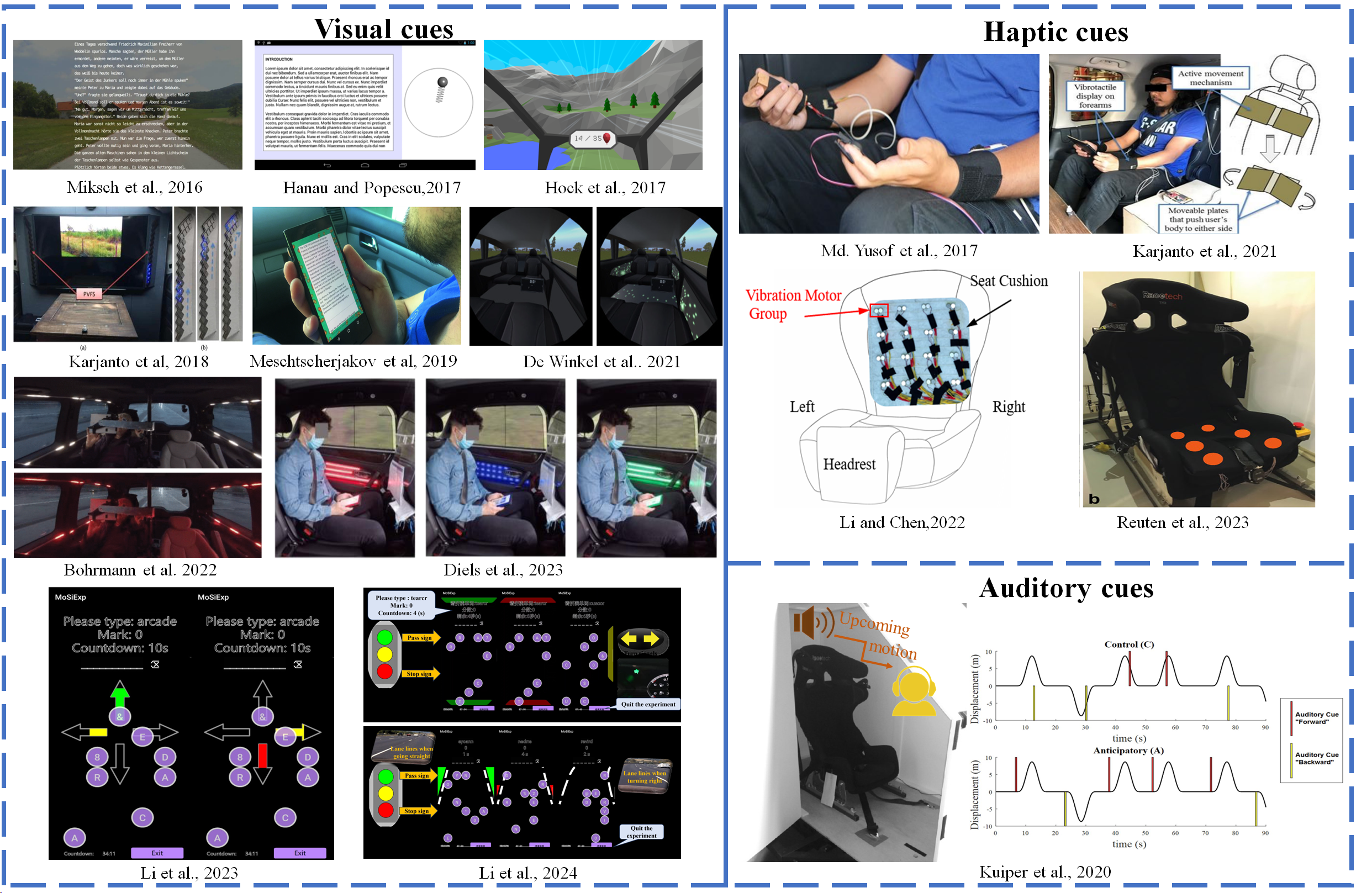}
	\caption{Example motion cues for MS mitigation}
	\label{fig_motion_cue_samples}
\end{figure*}

Table \ref{tab:table4} lists the various cueing methods developed in the literature, and 
Fig. \ref{fig_motion_cue_samples} offers some example design schemes.
It can be found that visual cueing is the most extensively studied approach, which often involves specific visual elements to provide passengers with motion information. 
In contrast, research on haptic and auditory cueing forms is still quite limited. 

\begin{table*}[htbp]
\caption{Motion cues for MS mitigation \label{tab:table4}}	\centering
\resizebox{\textwidth}{!}{
\begin{tabular}{lllll}
	\toprule
	Modality& Reference (author, year) & Experiment subjects & Analysis methods & Countermeasures\\
	\midrule
	
	\multirow{30}{*}{Visual}&\multirow{3}{*}{Miksch et al., 2016\cite{Miksch_2016_motionsicknessprevention}}&\multirow{2}{*}{12 persons (7 males, 5 females)}&\multirow{2}{*}{SSQ}&Real-time\\
	&&\multirow{2}{*}{18-53 years old}&\multirow{2}{*}{A 1-10 points rating scale}&video stream of the road ahead\\
	&&&&as the reading background\\
	&&&&\\
	
	&\multirow{2}{*}{Hanau and Popescu, 2017\cite{Hanau_2017_motionreadervisualacceleration}}&26 persons (13 males, 13 females)&\multirow{2}{*}{MSAQ}&\multirow{2}{*}{Moving spring ball on screen}\\
	&&Average 25 years old& &\\
	&&&&\\
	
	&\multirow{2}{*}{Hock et al., 2017\cite{Hock_2017_carvrenablingincar}}&23 persons (18 males, 5 females)&\multirow{2}{*}{SSQ}&Matching vehicle motion\\
	&&Average 26.17 years old& &with visual input in VR\\
	&&&&\\
	
	&\multirow{2}{*}{Karjanto et al., 2018\cite{Karjanto_2018_effectperipheralvisual}}&20 persons (13 males, 7 females)&MSSQ, MSAQ&\multirow{2}{*}{A LED peripheral visual feedback system}\\
	&&Average 26.2 years old&MSDV, ECG&\\
	&&&&\\
	
	&\multirow{2}{*}{Meschtscherjakov et al., 2019\cite{Meschtscherjakov_2019_bubblemarginmotion}}&10 persons (4 males, 6 females)&\multirow{2}{*}{MSSQ, MSAQ}&Moving bubbles\\
	&&21-60 years old&&on the smartphone margin\\
	&&&&\\
	
	&\multirow{2}{*}{De Winkel et al., 2021\cite{DeWinkel_2021_efficacyaugmentedvisual}}&19 persons (7 males, 12 females)&MSSQ-short&\multirow{2}{*}{Moving light particles in VR}\\
	&&Average 27.7 years old&SSQ, FMS&\\
	&&&&\\
	
	&\multirow{2}{*}{Bohrmann et al., 2022\cite{Bohrmann_2022_effectsdynamicvisual}}&23 persons (14 males, 9 females)&MSSQ&\multirow{2}{*}{A LED peripheral visual feedback system}\\
	&&Average 42.57 years old&MSAQ, FMS&\\
	&&&&\\
	
	&\multirow{2}{*}{Cho and Kim, 2022\cite{Cho_2022_ridevrreducingsickness}}&15 persons (14 males, 1 females)&\multirow{2}{*}{SSQ}&Mixed-in presentation of\\
	&&Average 24.67 years old& &motion flow information in VR\\
	&&&&\\
	
	&\multirow{2}{*}{Li et al., 2023\cite{Li_2023_mosiappmotion}}&26 persons (13 males, 13 females)&\multirow{2}{*}{MSSQ, SSQ, MISC}&\multirow{2}{*}{Arrows filled with colors on screen}\\
	&&Average 22.73 years old&&\\
	&&&&\\
	
	&\multirow{2}{*}{Diels et al., 2023\cite{Diels_2023_designstrategiesalleviate}}&16 persons (8 males, 8 females)&MISC&\multirow{2}{*}{A LED display with colors}\\
	&&30-60 years old&MSSQ-short&\\\midrule
	
	\multirow{14}{*}{Haptic}&\multirow{2}{*}{Md. Yusof et al., 2017\cite{Md.Yusof_2017_experimentalsetupmotion}}&10 persons (4 males, 6 females)&MSSQ, MSAQ&A stretchable fabric\\
	&&18-36 years old&MSDV&with vibration motors on forearms\\
	&&&&\\
	
	&\multirow{2}{*}{Md. Yusof et al., 2020\cite{Md.Yusof_2020_gainingsituationawareness}}&20 persons (12 males, 8 females)&MSSQ, MSAQ&A stretchable fabric\\
	&&18-47 years old&MSDV&with vibration motors on forearms\\
	&&&&\\
	
	&\multirow{2}{*}{Karjanto et al., 2021\cite{Karjanto_2021_onroadstudymitigating}}&18 persons (9 males, 9 females)&MSSQ-short, MSAQ&Vibration motors on forearms\\
	&&Average 28.4 years old&MSDV&Pushing passengers toward the turn direction\\
	&&&&\\
	
	&\multirow{2}{*}{Li and Chen, 2022\cite{Li_2022_mitigatingmotionsickness}}&20 persons (19 males, 1 female)&MSSQ-short&A seat cushion\\
	&&Average 23.1 years old&ECG, MISC&with vibration motors array\\
	&&&&\\
	
	&\multirow{2}{*}{Reuten et al., 2023\cite{Reuten_2023_effectivenessanticipatoryvibrotactile}}&20 persons (3 males, 17 females)&MISC&A seat cushion\\
	&&Average 26 years old&MSSQ-short&with vibration motors array\\
	\midrule
	
	\multirow{8}{*}{Auditory}&\multirow{2}{*}{Kuiper et al., 2020a\cite{Kuiper_2020_knowingwhatcoming}}&20 persons (12 males, 8 females)&MSSQ-short&\multirow{2}{*}{Anticipatory audio cues}\\
	&&Average 39.47 years old&MISC&\\
	&&&&\\
	
	&\multirow{2}{*}{Kuiper et al., 2020b\cite{Kuiper_2020_knowingwhatcominga}}&17 persons (5 males, 12 females)&MSSQ-short&\multirow{2}{*}{Anticipatory audio cues}\\
	&&Average 39.64 years old&MISC&\\
	&&&&\\
	
	&\multirow{2}{*}{Gálvez-García et al., 2020\cite{Galvez-Garcia_2020_decreasingmotionsickness}}&48 persons (22 males, 26 females)&MSSQ, SSQ&\multirow{2}{*}{Auditory stimulation}\\
	&&Average 21.58 years old&Head sway&\\
	\bottomrule
\end{tabular}}
\end{table*}

\subsubsection{(1) Visual Cues}
Motion cues based on visual interaction are the most popular. As the most important device for NDT, smartphones have been extensively explored to provide visual motion cues to users. For instance, a real-time video stream of the road ahead can be shown as the screen background when passengers are using their smartphones for reading, which can allow passengers to obtain information about the road ahead without affecting reading experience \cite{Miksch_2016_motionsicknessprevention}. 
It is also feasible to abstract some animation cues displayed on the screen, which are associated with vehicle motion, such as a rolling spring ball, floating bubbles, color-coded arrows, dashed lane lines, etc. \cite{Hanau_2017_motionreadervisualacceleration,Meschtscherjakov_2019_bubblemarginmotion, Li_2023_mosiappmotion,Li_2024_canweusesmartphone}. It is worth noting that animation cues should be perceived by the observer while avoiding occupying excessive visual attention. Therefore, they are often designed as moderate as possible and arranged in the periphery of visual field.

A slight drawback of most visual cues is that they only provide real-time or approximately real-time hint about vehicle motion to passengers. However, there is evidence suggesting beneficial effects of anticipating upcoming motion on MS mitigation \cite{Griffin_2004_visualfieldeffects,Wada_2018_analysisdriverhead}. 
Anticipatory motion cues have already been validated in simulation, but as autonomous driving matures, it becomes possible to embed the future motion in cue contents. 
Since human's peripheral vision is found to be sensitive to illumination and motion, or more specifically the optical flow, whereas our central vision is better at identifying some object. Based on this, a peripheral visual feedforward system was designed with two displays consisting of LED strips on the left and right, which could indicate upcoming motion by the LEDs light flow in corresponding direction \cite{Karjanto_2018_effectperipheralvisual}. 
Similar design can be migrated to ambient lights in vehicle, for which certain ways of lighting can inform passengers of upcoming motion, such as one certain color or flow direction \cite{Bohrmann_2022_effectsdynamicvisual,Diels_2023_designstrategiesalleviate}.

Recently, VR gaming is also becoming popular during car rides. However, using VR device may exacerbate MS in a moving vehicle. According to sensory conflict theory, the main reason for such MS is the mismatch between VR motion states perceived by visual receptors and that perceived by vestibular organs. Therefore, it is a feasible way to align users' visual and vestibular perceptions to mitigate sensory conflict in using VR in a moving car. This means that a mapping should be established between vehicle motion and VR motion in order to compensate for the impact of vehicle motion on the VR system, so that vehicle acceleration is in line with that in VR environment \cite{Hock_2017_carvrenablingincar,Cho_2022_ridevrreducingsickness}. With a similar idea, a rowing game for passengers is designed with its scenario mapped from the real driving scenario, while the alignment of virtual gaming contexts with real vehicle motion can alleviating MS \cite{ZhejiangUniversity_2020_methodsimulatingrowing}.

However, the benefits achieved solely through visual cues may have a ceiling effect of MS alleviation \cite{DeWinkel_2021_efficacyaugmentedvisual}. In such cases, visual cues may occupy too much visual attention, suggesting that ergonomic optimization is necessary.

\subsubsection{(2) Auditory Cues}
Researches show that if passengers receive auditory cues about the time and direction of upcoming movements, they can gain situation awareness and average MS ratings will decline significantly \cite{Kuiper_2020_knowingwhatcoming,Kuiper_2020_knowingwhatcominga}. 
Even more surprisingly, simply playing white noise in occupants' headphones can have a positive effect on body balance, thereby enhancing passengers postural stability \cite{Galvez-Garcia_2020_decreasingmotionsickness}.

There are several unique advantages with auditory interaction. One advantage is that sound can travel freely, especially in the enclosed vehicle cockpit. When there are multiple occupants in the vehicle, motion cues can be transmitted to everyone through auditory interaction. The fact that it occupies very little attention resource is another advantage. According to multiple resource theory \cite{Wickens_2008_multipleresourcesmental}, human brain can process auditory and visual information simultaneously. Thus, the visual NDT would be minimally affected with auditory cues. Certainly frequent auditory cues may be detrimental when auditory attention resource is scarce.

\subsubsection{(3) Haptic Cues}
Comparing to visual and auditory cues, haptic interaction for cueing has not been explored enough. The few existing studies have been conducted to prompt passengers by adding extra haptic interactive devices, like a stretchable fabric with several micro-vibration motors attached to the passenger's forearm \cite{Md.Yusof_2017_experimentalsetupmotion}. When the vehicle is about to turn left or right, the corresponding micro motors on the left or right forearm will vibrate to indicate the motion direction. Preliminary experiments showed positive effects but further experiments on the real roads did not support it \cite{Md.Yusof_2020_gainingsituationawareness}. 
Although passengers indeed reported a gain in situation awareness after perceiving haptic feedback, it did not help to mitigate MS. One reason for this result may be that the haptic feedback at user's forearms is unable to give more specific and detailed information such as cornering speed and curvature, etc., other than the next turning direction. Consequently it is still difficult for passengers to make appropriate adjustments to the vehicle's movement, even though they are mentally prepared. In addition to the vibrating device, Karjanto et al. \cite{Karjanto_2021_onroadstudymitigating} further added a device that would actively push passenger's shoulder toward the turn direction, which showed significant MS mitigation effect.

Further, a seat cushion with a vibration motor array was designed to provide passengers with haptic feedback containing more specific motion information via vibration waves of different amplitudes and frequencies \cite{Li_2022_mitigatingmotionsickness}. In this research, haptic cues that contain motion information could not only help most users to accurately predict the coming vehicle motion, but can also alleviate MS. 
Nevertheless, a recent research based on linear sled experiments did not reveal effectiveness of haptic cues, though the participants thought cues were helpful \cite{Reuten_2023_effectivenessanticipatoryvibrotactile}.

The best merit of haptic interaction is that it is covert and moderate, thus less invasive for NDT that usually require visual and auditory attention. On the other hand, since it is less intuitive in comprehending motion cues, the high cost of cognition and training is one drawback. 

\subsubsection{(4) Summary}
Presenting motion cue to passengers has become one of the most popular research directions in MS alleviation. One main reason is that these cue-based approaches are especially implementable in smart cockpit of intelligent vehicles, since there are widely available interaction modalities for cue delivery. Their lower cost comparing to that of automated planning and control approaches is another advantage, meaning that they can be used even for conventional vehicles.

Surely, ergonomics plays important roles in designing motion cues, since the cueing efficacy needs to be guaranteed without too much cognition resources. Although most of the literature present their design motivations and logic, here we summarize the key aspects in designing effective and acceptable cues for MS mitigation.
\begin{enumerate}
	\item \textbf{Frequency of cue delivery}. Finding a moderate cue frequency in time is a matter of deliberation. Too sparse or too frequent cue will damage its effect of MS alleviation. Too sparse cues can not help users gain sufficient situation awareness or anticipation, while too frequent cues are irritating, which would result in terrible user acceptance. 
	
	\item \textbf{Magnitude of cueing doses}. Similarly, if the magnitude of cues is designed to be too large, it may be too invasive to passenger NDT. On the contrary, it would also prevent cues from working if the magnitude is designed too mild to be noticeable. 
	
	\item \textbf{Onset and offset timing of cue}. The timing of cues should be aligned with the motion characteristics. If cues occur inappropriately in time, passengers get wrong motion information and sensory conflict may be intensified instead, which leads to more severe MS. Moreover, due to individual differences in reaction time, the timing of cues should be adjusted appropriately.
	
	\item \textbf{Scenario contexts}. It is crucial to choose the most suitable interaction modality according to the scenario contexts of passenger behaviors. For instance, when passengers are reading or watching a movie, visual cueing is not a good choice because they may take over visual attention resources. Clamorous cueing audios may cause unbearable experience when passengers are listening to music or talking. Further, the detailed strategy of cue delivery should also be optimized for specific scenarios.	

\end{enumerate} 

Another noteworthy issue is that almost any cue design needs user training, but an intuitive design with appropriate metaphors can contribute to user convenience.
Considering the personal invariability in using cues, user acceptance issues are still challenging. 
With the rapid development of smart cockpit technologies, perhaps it is a satisfactory approach to let passengers themselves customize the details of cues, including frequency, magnitude, timing and interaction mode, based on their preferences.

\section{Limitations, challenges and opportunities}
\label{sec:discuss}
\subsection{Limitations and challenges}

\begin{enumerate}
\item \textbf{MS Mechanism model}. \\
Ideally, a top-down approach to solving MS issues is to start from fundamental mechanisms, while unfortunately current MS models are still far from adequate. 
Particularly, the corresponding supporters for the two most popular theories, i.e. sensory conflict and postural instability theories, have been debating about their models' reasonability for decades. 

One fortunate thing is that their derived models have been well adopted in designing MS solutions. For example, an SVC model of MS is directly embedded in cruise control algorithm \cite{Li_2021_automatedcarfollowingalgorithm}. 
While for motion cue designing, both models can be used to generate basic principles of promoting motion anticipation for passengers \cite{Griffin_2004_visualfieldeffects,Li_2023_mosiappmotion}. 
However, due to unclear mechanism of MS, it is still difficult to apply quantitative models in optimizing MS countermeasures, especially when considering the complex scenarios in practice.  
The next-gen MS models may be available soon with the support of machine learning.

\item \textbf{MS quantification and evaluation}. 
\begin{itemize}
	\item For quantifying MS status, subjective questionnaires (e.g. MISC) are still the most dominant, while different questionnaires have different shortcomings in comprehensiveness, time to complete, etc.  
	\item For objective MS quantification, it is still difficult to establish a reliable mapping between MS status and objective indexes. Clearly, a single physiological indicator is not sufficiently accurate for various contexts, so perhaps fusioning multiple indexes corresponding to various MS is more trustworthy 
	\cite{Tan_2022_motionsicknessdetection,Hwang_2022_classificationmotionsickness}.
	\item It is difficult to guarantee the effectiveness of current MS quantification approaches for on-board usage, especially due to individual variability and environment robustness issues. 
	\item There still lacks a standard of MS quantification or evaluation, which makes it difficult to carry out parallel comparisons between different works.
\end{itemize} 

\item \textbf{MS countermeasures}. \\
There are already plenty of countermeasures to alleviate MS, yet effective and optimized countermeasure application are still on the way.
\begin{itemize}
	\item Existing countermeasures mostly focus on passengers' active adjustments and motion cues. Although they are cost-effective for application, they are merely a temporary fix rather than a complete cure for MS, since the uncomfortable vehicle motion as the root cause of MS is not directly suppressed or avoided. 
	\item MS mitigation via motion planning and control are emerging rapidly in the recent 5 years, however, much work needs to be done in this field. To name a few, these current approaches are not convincing enough, since there is still no open report of sickness-less planning or control algorithms that is validated via real road test.
	\item Most in-vehicle MS solutions require massive upgrades of original hardwares, meaning that the high cost may prevent them from being widely used. This is particularly true for those conventional vehicles with only limited driving assistance features, for which the applications of motion control are impossible.
\end{itemize}

\end{enumerate}

\subsection{Opportunities with intelligent vehicles}
Hopefully, intelligent vehicles are going to prevail in the near future, which will create many novel opportunities to MS alleviation.
\begin{enumerate}
	\item \textbf{Smooth ride: from planning to control}
	\begin{itemize}
		\item \textbf{Route planning}. With cloud computing and vehicle-to-everything (V2X) connectivity, the available information for route planning can be more comprehensive and updated, including the conditions of weather, road, traffic congestion, etc. Then similar to the manual way of trip planning, the time schedule, trajectory and even driving style of vehicle can be optimized for the least MS risk.   

		
		\item \textbf{Local motion planning}.
		Knowing traffic contexts in a broader scope of time and space, the vehicle longitudinal and lateral motion can be better planned for less MS. 
		For instance, when planning the trajectory for urban intersection driving, V2X can provide accurate information of the upcoming traffic signal phase and timing, then unnecessary sudden deceleration or lane change can be avoided, which is just like what experienced drivers do. 
		\item \textbf{Smooth control}. Vehicle motion control may be viewed as the last resort of MS mitigation before the real movement exerts on passengers.
		With advanced by-wire actuators of braking, traction and steering, smoothier motion can be achieved via optimizing MS indicators, e.g. MSDV.
		A recent progress on ride comfort control is the so-called predictive suspension control, which can perceive the road condition ahead and actively regulate suspension in advance to keep passengers stable. This is more implementable in intelligent vehicles, since they usually are equipped with cutting-edge lidar or high-resolution camera sensors.
	\end{itemize}

	\item \textbf{Smart cockpit: diverse interactions and motion cues}
	\begin{itemize}
		\item \textbf{Occupant status monitoring}. With multiple sensors of in-cockpit and wearable devices, it becomes more convenient to estimate passengers' states of discomfort, e.g. via facial expressions, body temperature, etc. Then based on these feedback, more appropriate countermeasures can be executed.
		\item \textbf{Comforting ride environment}. Miscellaneous cockpit environment control is possible to adjust temperature, humidity, ventilation, odor, quietness, seating, etc., which can be realized by regulating AC, window openning, and other smart components.
		\item \textbf{Customizable motion cues}. Thanks to the diverse interactive modalities, ergonomically-optimized motion cues can be well delivered to specific passengers, including their timing, magnitude, frequency, modality, etc., to help them reconstruct the situation awareness that is key to MS prevention. 
		Particularly, flexible switching among various modalities based on passengers' current NDT becomes possible.
	\end{itemize}

\item \textbf{Multi-solution integration and personalization}
\begin{itemize}
	\item \textbf{All-in-one integration}.
	Existing researches on alleviating MS often focus on a single mitigation countermeasure, but a combination of multiple countermeasures is now possible in intelligent vehicles. 
	\item \textbf{One-for-one personalization}.
	Due to individual variability among passengers, e.g. MS susceptibility and travel preferences, it is impossible to find one-size-fits-all solution for MS mitigation. 
	With passengers' digital-twin models from cloud, the intelligent vehicle can access the above individual appeals, and offer personalized cure of MS according to the realtime contexts of activities and scenarios, just like how doctors treating their patients. 
\end{itemize}
\end{enumerate}

\section{An example solution: integrated framework for motion sickness mitigation}
\label{sec:framework}

\begin{figure*}[htbp]
	\centering
	\includegraphics[width=1.0\textwidth]{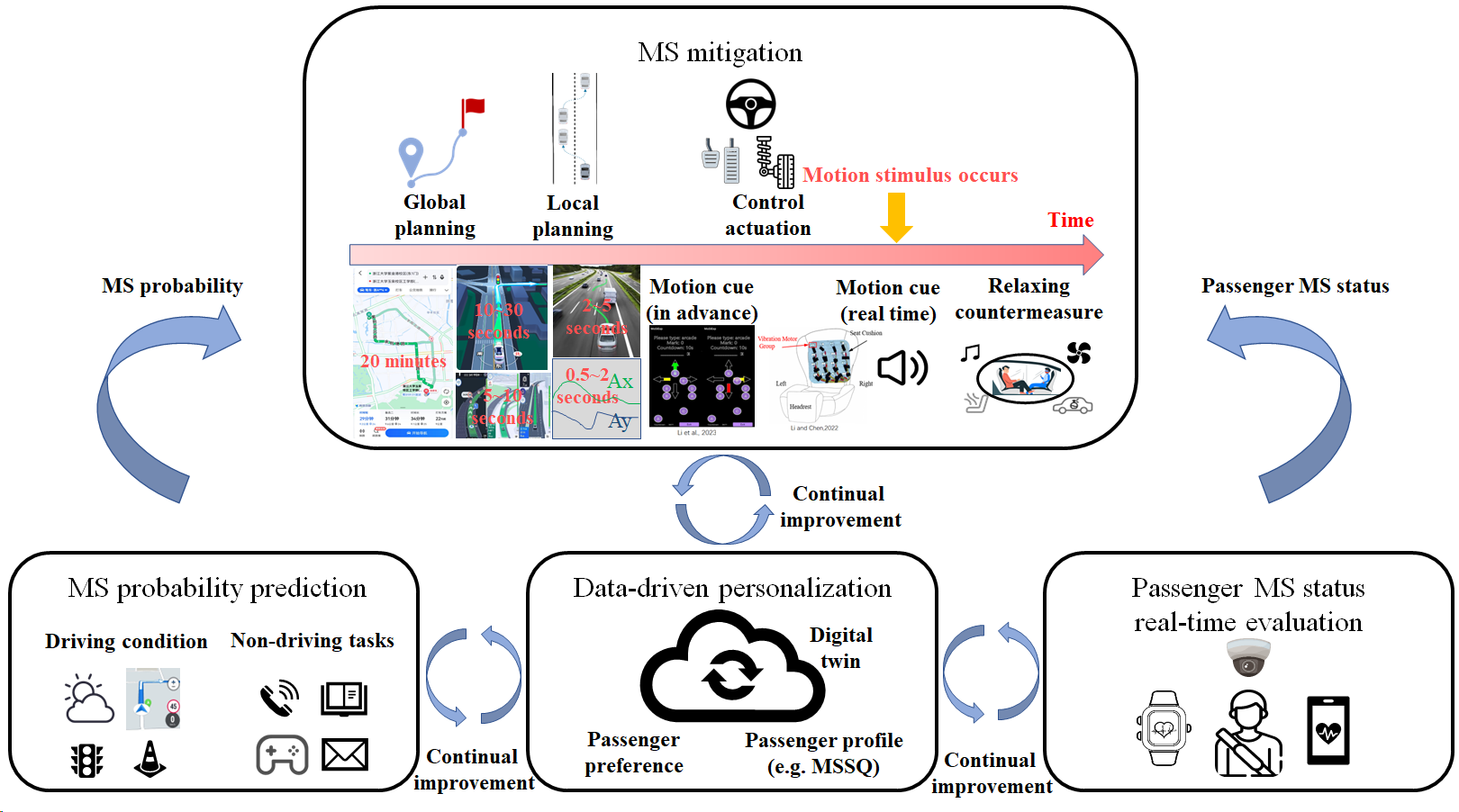}
	\caption{A carsickness mitigation framework for an intelligent vehicle.}
	\label{fig_framework}
\end{figure*}

The most impressive contribution of intelligent vehicles is that they make it possible to integrate various countermeasures into one synthesized solution.
Based on the above insights, we propose an integrated MS mitigation framework for intelligent vehicles, which covers the entire process of a trip, as shown in Fig.\ref{fig_framework}.

\subsection{Vehicle features}
The following assumptions of the vehicle are made here, for which Table \ref{tab:framework} lists the available vehicle features, including hardware, software and key system states.
\begin{itemize}
	\item The vehicle can work in the mode of automated driving, at least partially or conditionally automated, meaning that it has the ability to fulfill the SPPC driving tasks by its own. It is also equipped with predictive suspension control system. 
	\item The vehicle is equipped with occupant monitoring system, and the passenger MS status can be estimated based on non-invasive sensor information.
	\item The vehicle can be connected with a backend or cloud-based service platform to access or update the historical and current trip information, e.g. route plan, traffic prediction, real-time vehicle states, etc.
	\item The system can access the digital-twin information of users for personalization, including their preferences.
	\item The vehicle has a smart cockpit supporting active environment control and multi-modality motion cues. 
\end{itemize}
 
\begin{table*}[htbp]
	\caption{Available vehicle features for the example solution\label{tab:framework}}
	\centering
	\resizebox{\textwidth}{!}{
		\begin{tabular}{lllll}
			\toprule
			Modules & Sensors / actuators (hardware/software) & Sensed / regulated variables, states \\
			\midrule
			
			\multirow{5}{*}{Automated driving}
			&Driving style& Comfort, Mild, Sport, etc.\\
			&Route planning& Route info., includ. number of turns/ramp merging, etc.\\
			&Discrete decision&Yielding, preempting, lane change, uphill and downhill ride, etc.\\
			&{Path planning}&Short and long-term planning results includ. station, heading, curvature, etc.\\
			&Speed planning&Planned vehicle speed profile\\
			
			\midrule
			
			\multirow{3}{*}{Environment perception module}&Driving scene&Rural roads, highways, urban road, etc.\\
			
			&Traffic flow&Congestion level, average speed, average traffic density\\
			
			&Road surface&Road material, roughness, slope\\\midrule
			
			\multirow{4}{*}{Vehicle dynamic controller}&Steering&Steering Angle\\
			
			&Accelerator pedal&Percentage of openning from 0 to 1\\
			&Brake pedal&Percentage of openning from 0 to 1\\
			&Preview suspension control&Ride mode/strategy, ride height, etc.\\
			\midrule

			\multirow{1}{*}{Global Navigation Satellite System}&Global positioning \& IMU&Vehicle velocities, accelerations\\\midrule
			
			\multirow{5}{*}{Smart cockpit}&Temperature and humidity sensors&Temperature, humidity, air flow rate\\
			
			&Window control &Percentage of openning from 0 to 1\\
			
			&Air conditioner (AC)&Fan speed, target temperature\\
			
			&Seat&Seat position, heating, ventilation, etc.\\
			&Occupant monitoring system& Head orientation, occupant visual attention area, NDT type, etc. \\
			
			\bottomrule
	\end{tabular}}
\end{table*}

\subsection{Solution framework}
Overall, this is a comprehensive framework for motion sickness solution that starts from long-term global planning, to short-term local trajectory planning, motion control, and finally cockpit adjustment. In addition, it also takes into consideration individual preferences and usage scenarios to provide highly customized solutions for passengers, aiming to alleviate or completely resolve passengers' motion sickness problems.


(1) \textbf{Global planning}. Based on ``motion sickness probability prediction'', the global route planning module will comprehensively predict the MS-proneness index of multiple alternatives, which is based on the traffic, road and weather conditions obtained from the cloud or V2X. 
And an optimal route less likely to evoke MS is recommended for passengers, along with available driving styles to choose.

(2) \textbf{Local planning}. 
Certainly, since the instantaneous traffic conditions are always changing, the planned route is more used as a reference base for short time domain local trajectory planning.
According to traffic light information and congestion level in a short time scale, the local trajectory planning module can provide a safe and efficient trajectory while eliminating sharp acceleration and turning to ensure smooth and gentle trajectory and reduce MS incidence.
For example, in Table \ref{tab:framework}, a proper decision of driving may be yielding or preempting at an intersection, or more detailed plan of vehicle station and heading angle during a lane change maneuver.

(3) \textbf{Control and actuation}.
Responding to the trajectory issued by the upper level, the control module will further moderate the vehicle's movement like a filter, and regulate the actuators to avoid unexpected aggressive action while meeting the requirements of tracking accuracy. 
The preview suspension system can react in advance according to the road conditions ahead, which can effectively cushion not only vertical motion caused by bumpy roads, but also horizontal motion caused by turning.

(4) \textbf{Smart cockpit}.
As the last ditch, the facilities in smart cockpit kick in, e.g. by adjusting ambient environment, delivering motion cues, etc. 

Upon the trip setoff, the ambient cabin environment is adjusted to be relaxing for less passenger MS. 
Along the trip, the occupant monitoring system can directly provide passenger head orientation, visual attention area and NDT type (as shown in Table \ref{tab:framework}), which can further help the module of ``passenger MS status realtime evaluation" to identify the MS status.

Once one passenger is detected as ``suffering from MS", some preferred countermeasures are automatically implemented. 
Depending on the passenger's motion sickness status and the contexts of NDT, these may include playing pleasant music, releasing pleasant fragrances, regulating seating posture, adjusting to comfortable temperatures via AC or window openning, etc.

Furthermore, motion cues are provided to passengers to help them gain traffic situation awareness through visual, auditory and/or haptic modalities. Thanks to the fusion of multiple solutions in one framework, the motion planning results of future motion can be timely handed over for cueing, helping passengers to prepare for the upcoming motion stimulation.
As illustrated in Fig. \ref{fig_framework}, depending on the planning horizon, different levels of motion cue advance time can be possible. For instance, based on the information from the local planning module, a sharp-turn cue is presented to passengers at 10 sec before the vehicle really starts to turn, reminding them to adjust their seating posture. While in an emergency scenario, a discrete decision from automated driving control algorithms indicates a necessary hard braking maneuver, then a hard braking cue can be provided at 1 sec before the vehicle actually brakes.  

(5) \textbf{Data-driven personalization}.
A passenger's preference and profile during a trip will be memorized and learned, generating a personal digital-twin in the platform for future ride services. On one hand, this will aid in determining the most suitable MS countermeasures and ambient settings for a specific passenger, e.g. ambient temperature, lighting, humidity, seating posture, background music, etc. On the other hand, the historical data of passenger MS status and the corresponding MS countermeasures implemented in a trip, will be further added to the platform for next-version evolution of MS alleviating solution. Particularly, the contribution of each countermeasure will be analyzed to offer insights in the software updating.

\section{Conclusion}
\label{sec:conclude}
In this comprehensive review, we offer a detailed summary of MS theories, quantification methods, and countermeasures. We also identify the current research's limitations and challenges, from which we present an example framework for MS mitigation by integrating the novel opportunities in intelligent vehicles.

It remains an open question how to fully leverage these opportunities in the era of intelligent vehicles. Further research is complex and requires collaborative efforts from multiple disciplines, such as physiology, psychology, human factors engineering, vehicle engineering, robotics, and automation.

Apart from the challenges outlined in \autoref{sec:discuss}, these collaborative efforts, while challenging, may at least encompass the following future work.
\begin{enumerate}
	\item \textbf{Open-source community and data sharing}. We advocate for the formation of a motion sickness research community that promotes the sharing or open-sourcing of various motion sickness experiments and countermeasures data. This initiative would establish a dataset hub for motion sickness, serving a function akin to ImageNet for the artificial intelligence research community. Specifically, it should encompass experiment data from individuals of diverse genders, ages, races, and nationalities.
	\item \textbf{Standard research protocol}. To facilitate effective communication among different research groups or entities, it is crucial to discuss and establish a standard research protocol. These standardized procedures should cover aspects ranging from motion sickness subject screening, experimental stimulation, motion sickness quantification, to a standard driving/ride cycle, similar to vehicle emission tests (e.g. the Worldwide harmonized Light duty Test Cycle, WLTC). Such cycle should include typical motion sickness-prone scenarios, such as frequent and aggressive lane changes, hard braking, and rapid acceleration. This standardization will enable horizontal comparison between different studies and accelerate the iterative process of research in overcoming motion sickness challenges.
	\item \textbf{Implementation in practice}. If the understanding of the motion sickness mechanism continues to pose challenges in the near future, then from an engineering standpoint, current motion sickness countermeasures should be integrated and validated in practice, particularly on intelligent vehicles. These engineering explorations will not only help clarify potential motion sickness mitigation approaches that can be commercialized but also contribute to the accumulation of experimental data, thereby advancing scientific research into motion sickness mechanisms.
\end{enumerate}


\begin{dci}
The author(s) declared no potential conflicts of interest with respect to the research, authorship, and/or publication of this article.
\end{dci}

\begin{funding}
This work was supported by the National Natural Science Foundation of
China under Grant 52372421 and the Department of Science and Technology of Zhejiang under Grants 2022C01241 and 2023C01238. 
\end{funding}

\bibliographystyle{SageV}
\bibliography{Ref.bib}

\setlength{\nomitemsep}{0.1cm} %
\printnomenclature[1.5cm] %

\begin{thenomenclature}
\nomgroup{A}
\item [{ECG}]\begingroup Electrocardiography \nomeqref {20}\nompageref{26}
\item [{EDA}]\begingroup Electrodermal activity \nomeqref {20}\nompageref{26}
\item [{EEG}]\begingroup Electroencephalography \nomeqref {20}\nompageref{26}
\item [{EGG}]\begingroup Electrogastrograph \nomeqref {20}\nompageref{26}
\item [{FMS}]\begingroup Fast motion sickness scale \nomeqref {20}\nompageref{26}
\item [{HR}]\begingroup Heart rate \nomeqref {20}\nompageref{26}
\item [{HRV}]\begingroup Heart rate variability \nomeqref {20}\nompageref{26}

\item [{MISC}]\begingroup Misery scale \nomeqref {20}\nompageref{26}
\item [{MPC}]\begingroup Model predictive control \nomeqref {20}\nompageref{26}
\item [{MS}]\begingroup Motion sickness \nomeqref {20}\nompageref{26}
\item [{MSAQ}]\begingroup Motion sickness assessment questionnaire \nomeqref {20}\nompageref{26}
\item [{MSDV}]\begingroup Motion sickness dose value \nomeqref {20}\nompageref{26}
\item [{MSI}]\begingroup Motion sickness incidence \nomeqref {20}\nompageref{26}
\item [{MSSQ}]\begingroup Motion sickness susceptibility questionnaire \nomeqref {20}\nompageref{26}

\item [{NDT}]\begingroup Non-driving related tasks \nomeqref {20}\nompageref{26}
\item [{NMPC}]\begingroup Nonlinear model predictive control \nomeqref {20}\nompageref{26}
\item [{OCP}]\begingroup Optimal control problem \nomeqref {20}\nompageref{26}
\item [{SPPC}]\begingroup Sensing - prediction - planning - control \nomeqref {20}\nompageref{26}
\item [{SSQ}]\begingroup Simulator sickness questionnaire \nomeqref {20}\nompageref{26}
\item [{SVC}]\begingroup Subjective vertical conflict \nomeqref {20}\nompageref{26}
\item [{VIMS}]\begingroup Visually induced motion sickness \nomeqref {20}\nompageref{26}
\item [{V2X}]\begingroup Vehicle to Everything \nomeqref {20}\nompageref{26}
\item [{WRMSA}]\begingroup Weighted root mean square acceleration \nomeqref {20}\nompageref{26}

\printnomenclature


\end{thenomenclature}

\end{document}